\documentclass[twocolumn]{revtex4}
\usepackage{mathrsfs}
\usepackage{amsfonts}
\usepackage{graphicx}
\usepackage{amsmath}
\usepackage{amssymb}
\usepackage{ifpdf}
\begin{document}
\title{Potential-driven eddy current in rippled graphene nanoribbons}
\author{Hua-Tong Yang}\email{yanght653@nenu.edu.cn}
\affiliation{Key Laboratory for UV-Emitting Materials and Technology
of Ministry of Education, Department of Physics, Northeast Normal
University, Changchun 130024, China}

\begin{abstract} It is well known that an eddy current will be induced in a conductor subject to a varying
magnetic field. Here we propose another mechanism of generating
nano-scale eddy current in rippled graphene nanoribbons(GNRs), which
is only driven by an electric potential. In particular, it is found
that under appropriate gate voltages, a local deformation may induce
some unexpected global eddy currents, which form vortices in both
rippled and entire flat areas of the GNR. We will explain that these
vortices in flat areas is a manifestation of the nonlocality of
quantum interference.
\end{abstract}
 \maketitle

\subsection{Introduction}Graphene, as the first experimentally
available strictly two-dimensional crystal\cite{Novoselov-sci04,
Novoselov-PNAS05}, has offered a lot of new possibilities for both
fundamental research and new technologies due to its exceptional
mechanical and electronic properties\cite{Novoselov-N05, ZhangY-05,
Neto-09}. These single-atom-thick carbon membranes are very soft and
flexible. Experiments have observed that graphene sheets are
rippling even in free-standing condition\cite{Meyer,Parga08,Bao09}.
This feature is expected to have significant influences on their
electronic properties and to result in some new observable
effects\cite{Juan-07,Guinea-09, Levy, Juan-11}. Recently, Bao et al.
reported a method to control and create nearly periodic ripples of
sinusoidal form in graphene sheets\cite{Bao09}, which provides a
possibility to design and manipulate electronic states in graphene
sheets by controlling their local structure. According to the
continuous model, a linear approximation of the tight-binding
Hamiltonian in vicinity of its Dirac points, the electronic states
of graphene can be described by a massless Dirac equation. In this
model the action of a ripple can be represented by an effective
vector field\cite{Suzuura-02,Manes-07,Vozmediano_10}, which
essentially is the shift of the Dirac points due to the deformation,
and a velocity tensor representing anisotropy of the energy
band\cite{Yang}. This continuous model is adequate if the
deformation is very smooth. However, this condition cannot be always
satisfied in realistic graphene sheets, since some corrugations in
graphene sheets are not so smooth.

In this work, the local density of states(LDOS) and current flow in
rippled GNRs are investigated by using the tight-binding model and
non-equilibrium Green's function(NEFG) method\cite{Kadanoff,
Keldysh, Chou}. It is found that the current distributions near each
step edge of the conductance staircases are unstable, they will
become non-potential or eddy flows if the GNRs are rippled, although
their driven force is potential. In particular, there also occur
some global eddy currents, which are still vortical in the flat
regions of the GNRs. The vortices in flat areas is rather exotic
from the classical view point, because there is no local deflection
mechanism in these areas. We will explain that this phenomenon is
originated from a non-local quantum interference effect. This eddy
current will give rise to a very inhomogeneous magnetic field in
nanometer scale. It not only has crucial influences on the
performance of graphene-based electronic devices and circuits, but
may also have important potential applications.

\subsection{Model and Method}The electronic properties of graphene
can be described by the nearest tight-binding Hamiltonian
\begin{eqnarray} \label{Hamilton}
\hat{H}=\sum_{<\mathbf{r},\mathbf{r'}>}t(\mathbf{r}-\mathbf{r'})
c^{\dag}(\mathbf{r})c(\mathbf{r'})+\textrm{H.c.}\end{eqnarray} where
$\mathbf{r},\mathbf{r'}$ denote two nearest lattice points,
$c^{\dag},c$ are electron's creation and annihilation operators,
$t(\mathbf{r}-\mathbf{r'})$ is a distance-dependent hopping
amplitude modeling the influence of deformation and can be fitted by
\begin{eqnarray}t(\mathbf{r}-\mathbf{r'})\simeq
t_0e^{\alpha(1-|\mathbf{r}-\mathbf{r'}|/l)},\end{eqnarray} where
$t_0=-2.75eV$, $\alpha\simeq 3.37$, $l\simeq
1.42{\AA}$\cite{Pereira, Pellegrino}. For the tight-binding model,
we define the bond current $j(\mathbf{r},\mathbf{r'})$ from
$\mathbf{r}$ to $\mathbf{r'}$ by the conservation equation
$\sum_{\mathbf{r'}} j(\mathbf{r},\mathbf{r'})=-\frac{\partial
\langle \hat{n}(\mathbf{r}t)\rangle}{\partial t}$, where
$\hat{n}(\mathbf{r}t)=2c^{\dag}(\mathbf{r}t)c(\mathbf{r'}t)$ is the
electronic number operator, the factor $2$ comes form the spin
degree of freedom. From the equation of motion we have
\begin{eqnarray}j(\mathbf{r},\mathbf{r'},t)=\frac{4}{\hbar}\textrm{Re}[t(\mathbf{r}-\mathbf{r'})G^{<}(\mathbf{r'}t,\mathbf{r}t)],\end{eqnarray}
where $G^{<}(\mathbf{r'}t',\mathbf{r}t)=i\langle
c^{\dag}(\mathbf{r}t)c(\mathbf{r'}t') \rangle $ is the lesser
Green's function. In steady states, the current can be written as
$j(\mathbf{r},\mathbf{r'})=\int j(\mathbf{r},\mathbf{r'},E)dE,$
where
\begin{eqnarray}j(\mathbf{r},\mathbf{r'},E)=\frac{4}{h}\textrm{Re}[t(\mathbf{r}-\mathbf{r'})G^{<}(\mathbf{r'},\mathbf{r},E)]\end{eqnarray}
is the current per unit energy\cite{Areshkin-07}.

\begin{figure}[htb]
\setlength{\unitlength}{1cm} \centering
\begin{picture}(8,3.5)(0,3.9)
\includegraphics[width=8cm]{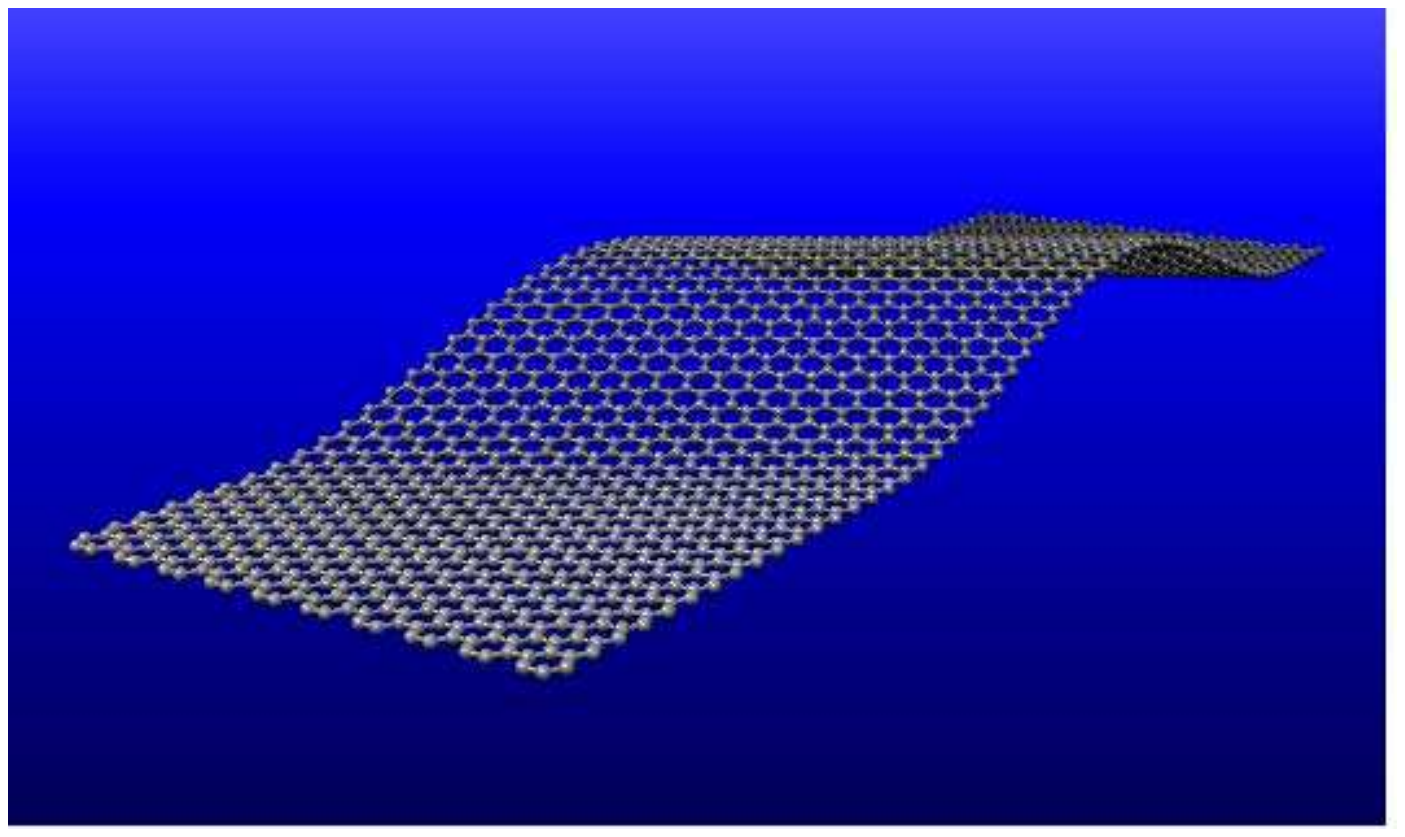}
\end{picture}
\caption{\label{ribbon} An $N=20$ zigzag GNR with a sinusoidal
ripple given by Eq.(\ref{sine}), here $h\simeq 2a$ with $a\simeq
2.46\AA$ the length of the basis vectors, the ripple wavelength
$\lambda=30a$, the included angle $\theta_{\mathbf{k}}$ between the
GNR axis and $\mathbf{k}$ is $30^{\circ}$, $\phi_0=-\pi/2$, $m=1$.}
\end{figure}

For the sake of simplicity, we only consider zigzag and armchair
GNRs with a finite rippled region. The ripple is a static height
fluctuation given by
\begin{eqnarray}\label{sine}z(\mathbf{r})=\left\{\begin{array}{ll}h\sin(\mathbf{k}\cdot \mathbf{r}+\phi_0),&0\leq\mathbf{k}\cdot\mathbf{r}\leq2m\pi\\
h\sin\phi_0,&\mathbf{k}\cdot\mathbf{r}<0~\textrm{or}
~\mathbf{k}\cdot\mathbf{r}>2m\pi,
\end{array}\right.\end{eqnarray} with $\phi_0$ an arbitrary constant and $m$ an integer, the atomic in-plane displacement is
ignored. A typical configuration of the rippled GNRs considered in
this paper is shown in Fig.\ref{ribbon}.

In order to calculate the LDOS and current of GNRs by using the NEFG
method\cite{Kadanoff, Keldysh, Chou}, the GNRs have to be considered
as consisting of three connected parts: a rippled region as a sample
and two semi-infinite ideal regions as
leads\cite{Lopez-Sancho,Lake}. The left and right leads can be
assumed to be in equilibrium states with different chemical
potentials $\mu_{L}$ and $\mu_R$, respectively, where $\mu_L-\mu_R$
is very small. Thus the current can be written as
\begin{eqnarray}j(\mathbf{r},\mathbf{r'})=\frac{4}{h}\int_{\mu_R}^{\mu_L}\textrm{Re}
[t(\mathbf{r}-\mathbf{r'})G^{<}(\mathbf{r'},\mathbf{r},E)]dE
,\end{eqnarray} where
\begin{subequations}\begin{align}&G^{<}=G^{r}\Sigma^{<}_{L}G^{a},\\
&G^{r,a}=(E^{\pm}I-H_c-\Sigma^{r,a})^{-1},\end{align}\end{subequations}
with $E^{\pm}\equiv E\pm i0^{+}$ and $H_c$ is the Hamiltonian matrix
of the isolated sample.
$\Sigma^{r,a}=\Sigma^{r,a}_{L}+\Sigma^{r,a}_{R}$ is the self-energy
arisen from the couplings between the sample and two leads.
$\Sigma^{<}_{L,R}=f_{L,R}(\Sigma^{a}_{R,L}-\Sigma^{r}_{R,L}),$ where
the Fermi functions $f_{L,R}$ satisfy
$f_{L,R}(E)\simeq\theta(\mu_{L,R}-E)$ at low temperature.
$\Sigma^{r,a}_{R,L}$ can be obtained by using the iterative method
developed by Lopez-Sancho et.al.\cite{Lopez-Sancho}. To reduce the
calculation and memory requirement, we orderly denote the layers of
the sample from the left to the right by $1,2,\cdots,l$, thus the
necessary sub-matrices $G^{<}_{n,n-1}$ and $G^{<}_{n,n}$ for
calculating current can be obtained by following back-and-forth
recurrence procedure:
\begin{subequations}\begin{align}&g^{r,a}_{n,n}=(E^{\pm}I-H_{n,n}-\Sigma^{r,a}_{R,n})^{-1},\\
&\Sigma^{r}_{R,n-1}=H_{n-1,n}g^{r}_{n,n}H_{n,n-1},\end{align}\end{subequations}
with $n=l,\cdots,2$ and $\Sigma^{r}_{R,l}=\Sigma^{r}_{R}.$ For
$\mu_R\leq E\leq \mu_L$,
\begin{subequations}\begin{align}&G^{<}_{n,n-1}=g^{r}_{n,n}H_{n,n-1}G^{<}_{n-1,n-1},\\
&G^{<}_{n,n}=G^{<}_{n,n-1}H_{n-1,n}g^{a}_{n,n}\end{align}\end{subequations}
with $n=2,\cdots,l$ and
$G^{<}_{1,1}=G^{r}_{1,1}\Sigma^{<}_{L}G^{a}_{1,1}$, where
\begin{eqnarray}G^{r,a}_{1,1}=[E^{\pm}I-H_{1,1}-\Sigma^{r,a}_{R,1}-\Sigma^{r,a}_{L,1})]^{-1}.\end{eqnarray}
Similarly, the LDOS \begin{eqnarray}\rho(\mathbf{r},E)
=\frac{1}{2\pi}A(\mathbf{r},\mathbf{r}),\end{eqnarray} where
$A=i(G^r-G^a)$, and the conductance
\begin{eqnarray}T(E)=\frac{2e^2}{h}\textrm{Tr}[\Gamma_R(A-G^{r}\Gamma_R
G^{a})],\end{eqnarray} with
$\Gamma_{R,L}=i(\Sigma^{r}_{R,L}-\Sigma^{a}_{R,L})$, can also be
obtained by the same method.

In order to see the ripple's influence on the current as clear as
possible, the inherent deflection due to the lattice background has
to be averaged out. To this end we define a cell-average current as
an average vector of six bond currents around a honeycomb cell
\begin{eqnarray}\mathbf{j}(\mathbf{r}_c)=\frac{1}{6}\sum_{i<j}
j(\mathbf{r}_i,\mathbf{r}_j)\frac{\mathbf{r}_j-\mathbf{r}_i}{|\mathbf{r}_j-\mathbf{r}_i|},\end{eqnarray}
where $\mathbf{r}_{i,j}$ denote the vertices of the cell,
$\mathbf{r}_c$ is the cell center.

\subsection{Results and Discussion}
\begin{figure}[htb]
\setlength{\unitlength}{1cm}
\centering
\begin{picture}(8,4.7)(0,0.7)
\includegraphics[width=8cm]{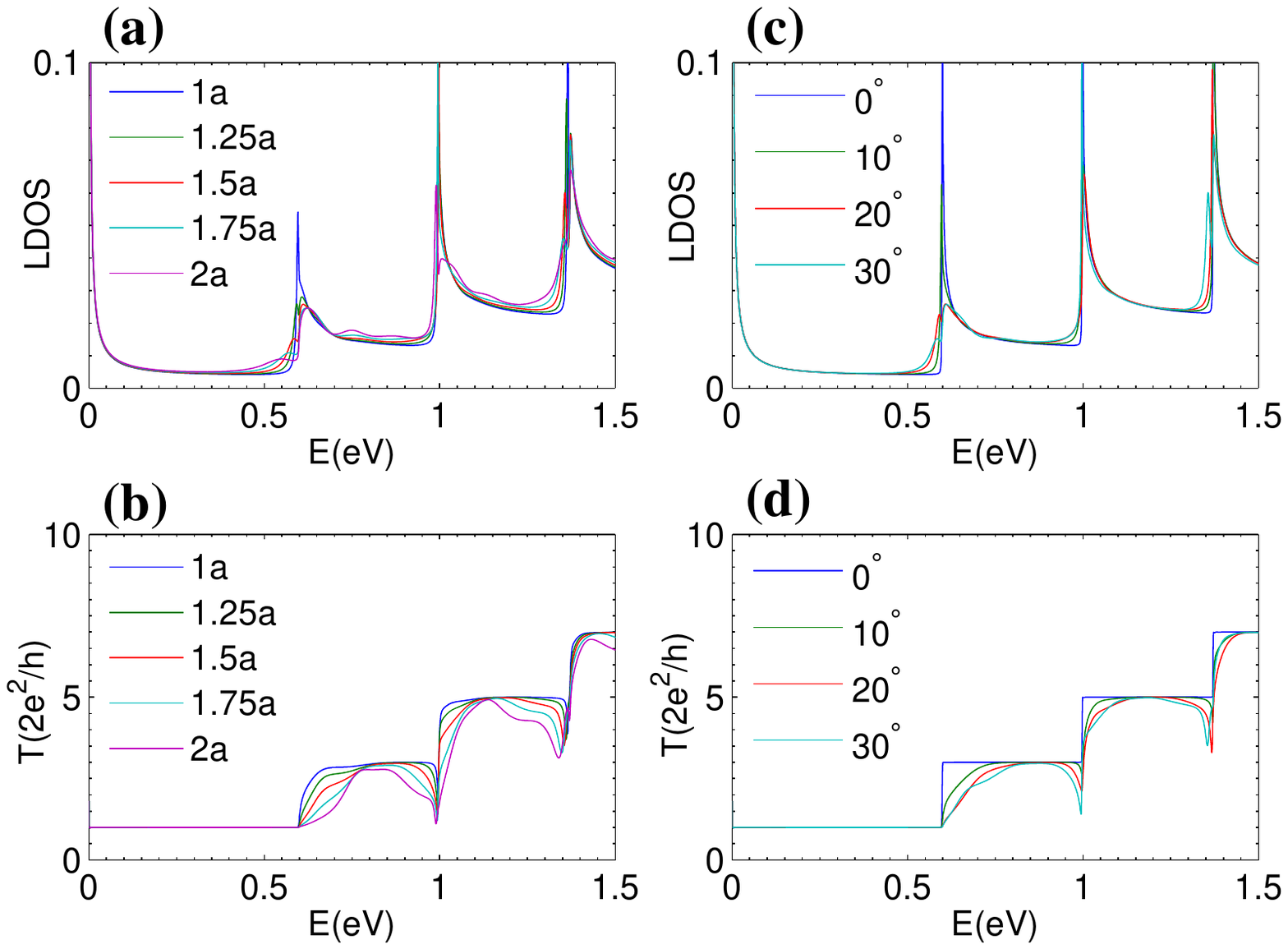}
\end{picture}
\begin{picture}(8,4.7)(0,0.7)
\includegraphics[width=8cm]{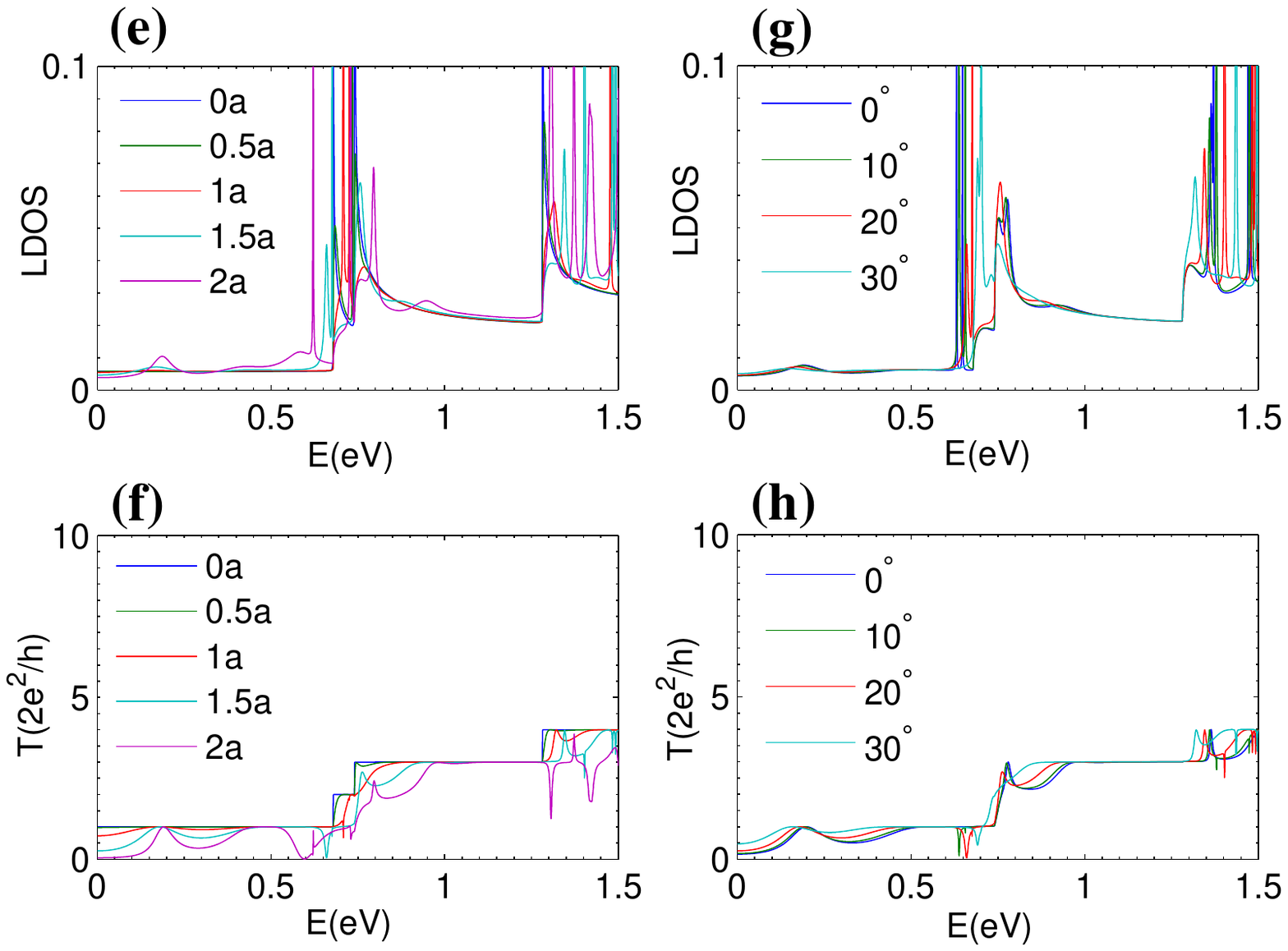}
\end{picture}
\caption{\label{LDOS} (a) LDOS and (b) conductance of rippled $N=20$
zigzag GNRs as shown in Fig.\ref{ribbon} for different $h$'s, where
$\lambda=30a$, $\theta_{\mathbf{k}}=30^{\circ}$. (c) LDOS and (d)
conductance of identical GNRs as Fig.\ref{ribbon} for different
$\theta_\mathbf{k}$'s, where $h=1.5a$, $\lambda=30a$. (e) LDOS and
(f) conductance of $N=20$ armchair GNRs for different $h$'s, here
$\lambda=40a$, $\theta_\mathbf{k}=20^{\circ}$. (g) LDOS and (h)
conductance of identical GNRs as (e,f) for different
$\theta_\mathbf{k}$'s, where $h=1.5a$, $\phi_0=-\pi/2$.} 
\end{figure}

First we compare the LDOSs and
conductances for a fixed GNR with different ripples. Fig.\ref{LDOS}
shows the LDOSs and conductances for a zigzag and an armchair GNR in
the presence of different ripples given by Eq.(\ref{sine}) with
different $h$'s and orientations (represented by slope angle
$\theta_\mathbf{k}$).
The most remarkable change is that the conductance near every step
edge of the conductance staircase is remarkably decreased.
Meanwhile, each corresponding van Hove peak in LDOSs is also broaden
and successively split into two sub-peaks when $h$ reaches critical
values, except for zigzag GNRs with
$\theta_\mathbf{k}=0^{\circ}$(which will be further discussed in the
last paragraph). The critical degrees of deformation can be
represented by the associated maximum relative bond elongation due
to the ripple. This bond elongation can be roughly estimated from
the ratio between the ripple's height $h$ and wavelength $\lambda$
if we consider only the atomic height fluctuation. For zigzag GNRs
with $\theta_\mathbf{k}=30^{\circ}$, this critical ratio is about
$1/20$, the corresponding maximum relative bond elongation is about
$5\%$.

\begin{figure}[htb]
\setlength{\unitlength}{1cm}
\centering
\begin{picture}(8.0, 5.3)(0, 0.3)
\includegraphics*[width=8cm]{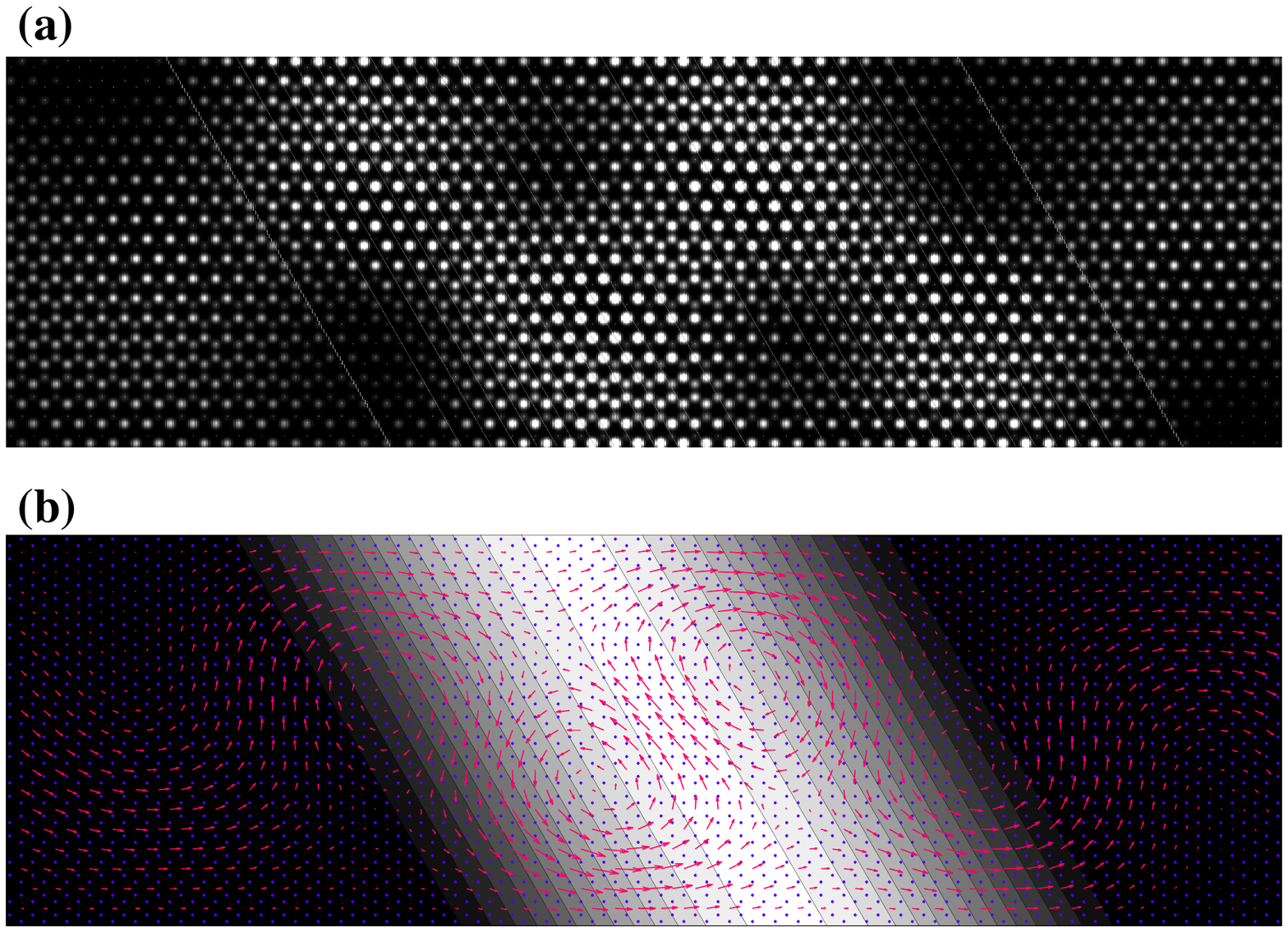}
\end{picture}
\begin{picture}(8.0, 5.3)(0, 0.3)
\includegraphics[width=8cm,clip]{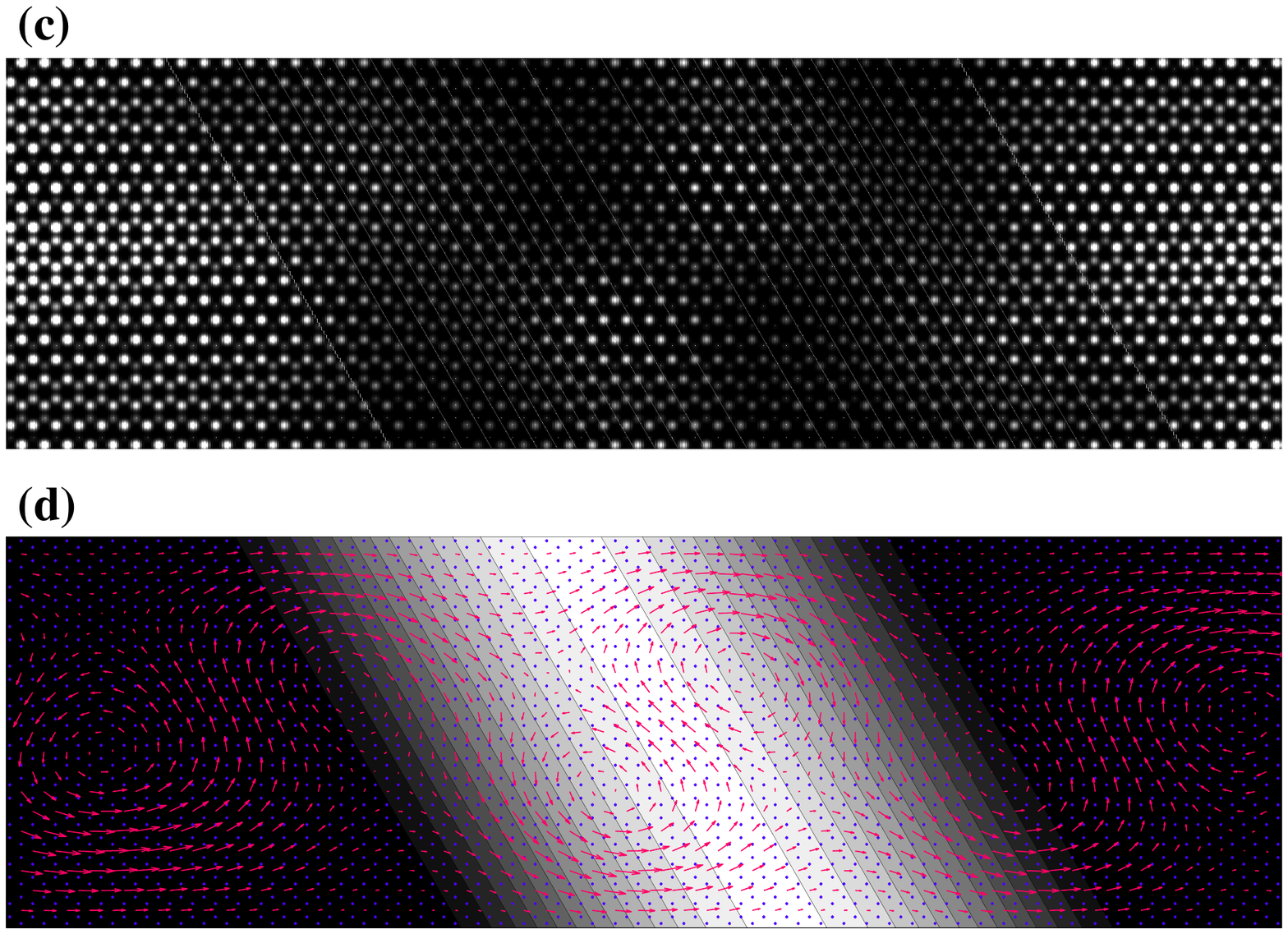}
\end{picture}
\caption{\label{E56-61} (a) LDOS and (b) current distribution at
$0.56eV$(the lower sub-peak split off from 1st van Hove peak) in an
$N=20$ zigzag GNR as shown in Fig.\ref{ribbon}. (c) LDOS and (d)
current at $0.61eV$(the upper sub-peak of the 1st van Hove peak) in
the same GNR.  The brightness in (a,c) represents the LDOS, the
contours depict the ripple. The brightness in (b,d) represents the
height of the ripple.}
\end{figure}

To get an intuitive understanding of the microscopic behavior of
electrons that gives rise to these changes, now we analyze the
spatial distributions of the LDOS $\rho(\mathbf{r},E)$ and current
$\mathbf{j}(\mathbf{r}_c,E)$ at the corresponding energies.
Fig.\ref{E56-61} shows two examples of this kind of LDOS and current
distributions in a rippled $N=20$ zigzag GNR at $0.56eV$ and
$0.61eV$, respectively corresponding to two sub-peaks split off from
the first van Hove peak in conducting sub-band. We can see from the
LDOS distributions (Fig.\ref{E56-61}(a,c)) that the lower sub-peak
is mainly localized in the ripple area; while the upper one
(belonging to the bottom of the first conducting sub-band) is
extended, its LDOS does not decay outside the ripple region. More
importantly, their current distributions both occur remarkable
vortices (Fig.\ref{E56-61}(b,d)). Some current lines even form
closed loops, showing that these currents are not potential flow.
Similarly, the second van Hove peak also splits into two sub-peaks
and their current distributions also have this vortical feature. The
valence bands also occur identical phenomena. When the bias is
reversed, these eddy currents will also be exactly reversed, thus an
alternative bias can drive a varying eddy current in this kind of
rippled GNRs.

At first sight, these vortices seems can be ascribed to Landau-like
quasi-bound states caused by the ripple-induced pseudo-magnetic
field, since according to the continuous model a geometrical
deformation will induce an effective vector potential
\begin{eqnarray}\label{potential}\mathbf{A}\simeq\frac{1}{2}(\sqrt{3}(t_3-t_2),t_3+t_2-2t_1)\end{eqnarray}
with $t_{1,2,3}$ three nearest hopping
amplitudes\cite{Vozmediano_10}. Actually, some conductance
fluctuation phenomena similar to the results shown in
Fig.(\ref{LDOS}) have been observed in the measurement of
differential conductance of graphene sheet with nanobubbles, and the
researchers interpreted all this conductance fluctuation as the
contribution of
 the Landau levels arising from the pseudo-magnetic field\cite{Levy}. It
is true that most of the above results can be explained by this
pseudo-magnetic field. For example, if $\mathbf{k}$ is exactly along
some special directions, e.g.,
$\theta_\mathbf{k}=0^{\circ},30^{\circ},60^{\circ}$, we can deduce
some general properties of $\mathbf{A}$ by qualitative analysis,
which can be compared with the numerical results. In these special
cases the $\mathbf{A}$ will has the form
$\mathbf{A}_{0}\cos(\mathbf{k}\cdot \mathbf{r}+\phi_0)$ and will be
parallel or perpendicular to $\mathbf{k}$\cite{Yang}. More
specifically, for rippled zigzag GNRs of
$\theta_\mathbf{k}=0^{\circ},60^{\circ}$ or armchair GNRs of
$\theta_\mathbf{k}=30^{\circ}$, $\mathbf{A}$ is nearly parallel to
$\mathbf{k}$, so the pseudo-magnetic field
$\nabla\times\mathbf{A}\simeq-\mathbf{k}\times\mathbf{A}_0\sin(\mathbf{k}\cdot
\mathbf{r}+\phi_0)$ is very small, thus the ripple influence will be
much weaker than other ripple orientations; while for armchair GNRs
of $\theta_\mathbf{k}=0^{\circ}$ or zigzag GNRs of
$\theta_\mathbf{k}=30^{\circ}$, $\mathbf{A}$ is perpendicular to
$\mathbf{k}$, so the pseudo-magnetic field is strong and will
strongly disturb the electronic states. For these special cases, the
above explanation is qualitatively in accord with the result in
Fig.\ref{LDOS}(there are still some more subtle problems, which will
be discussed in last paragraph).

However, these current distributions also exhibit a remarkable
feature revealing that this is not the whole story. We notice that
if the energy slightly higher than the bottom of a conducting
sub-band (or lower than the top of a valance sub-band), the vortices
of the eddy current will not entirely lie within the ripple region,
but also occur in flat regions far from the ripple, where the
pseudo-magnetic field has vanished, as can be seen from
Fig.\ref{E56-61}(d). This feature reveals that these vortices cannot
be interpreted as Landau states produced by the pseudo-magnetic
field, because all ripple-induced scattering mechanisms, both the
pseudo-magnetic field and the velocity variation, only act locally
within the ripple area, hence the electronic waves should freely
propagate in flat areas.
\begin{figure}[htb]
\setlength{\unitlength}{1cm}
\centering
\begin{picture}(8, 5.2)(0,0.5)
\includegraphics[width=8cm,clip]{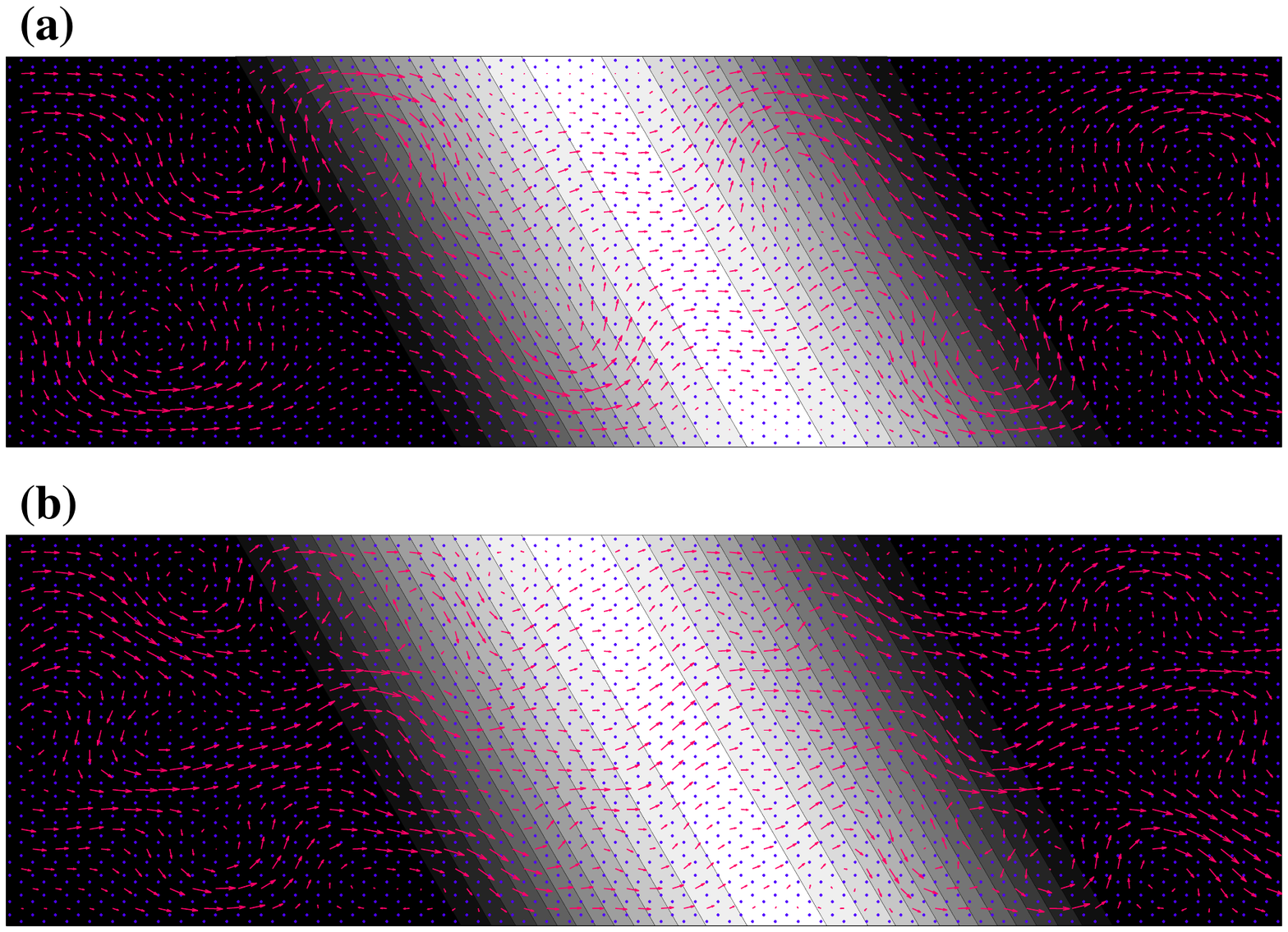}
\end{picture}
\begin{picture}(8, 5.2)(0,0.5)
\includegraphics[width=8cm,clip]{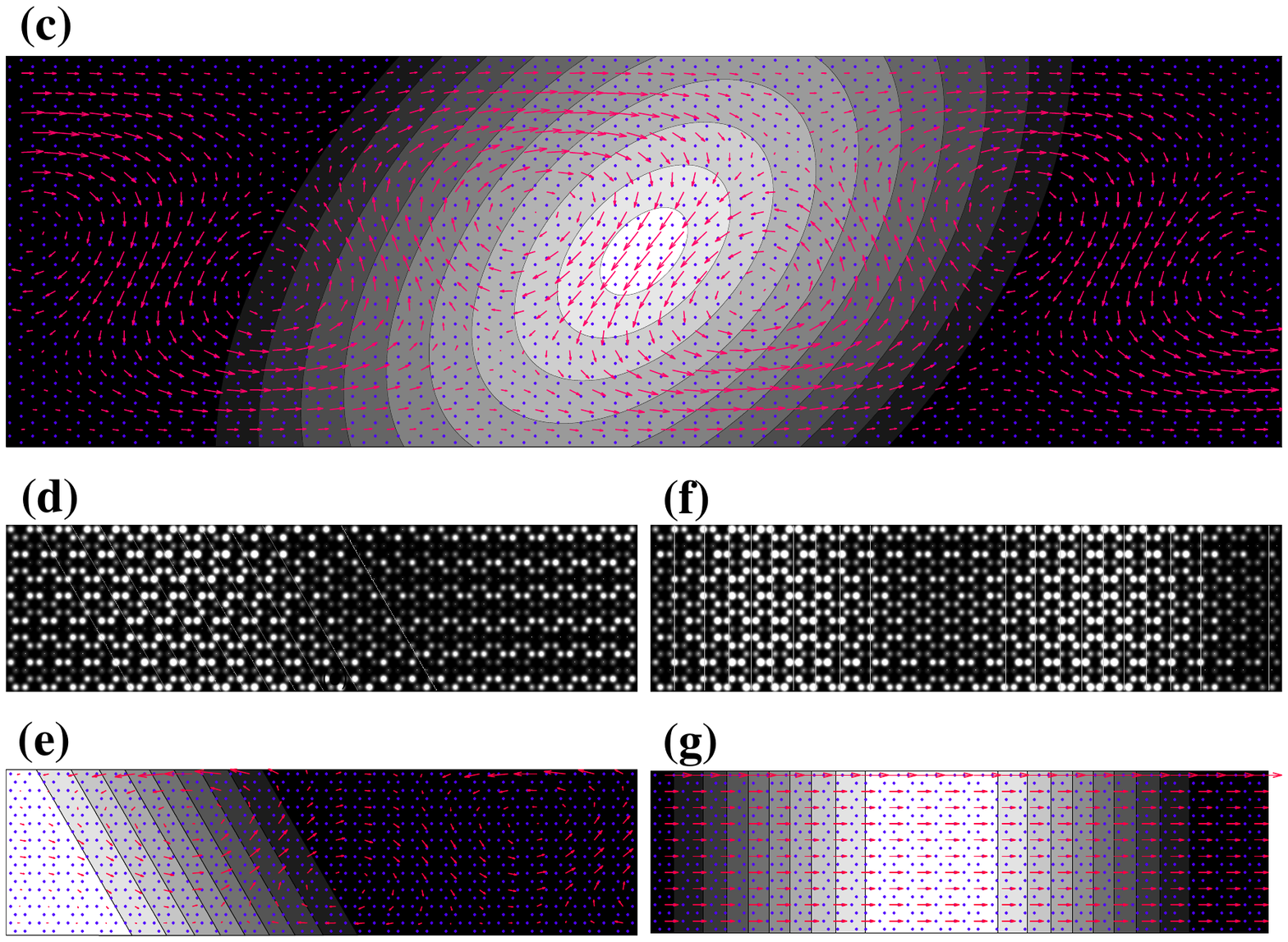}
\end{picture}
\caption{\label{eddycurrent} (a) Eddy current in the same GNR as
Fig.\ref{ribbon} at $1eV$, the upper sub-peak of the second van Hove
peak, and (b) $1.37eV$, the upper sub-peak of the third van Hove
peak. (c) Eddy current at $0.6eV$ in a similar GNR as
Fig.\ref{E56-61}(c,d) but with a hyperbolic-surface ripple given by
Eq.(\ref{hypersurf}). (d) LDOS and (e) current distribution at
$0.69eV$ in an $N=20$ armchair GNR with a ripple given by
Eq.(\ref{sine}), where $h=1.5a$, $\lambda=40a$,
$\theta_\mathbf{k}=30^{\circ}$, $\phi_0=-\pi/2$. (f) LDOS and (g)
current distribution at $0.631eV$ in a GNR as (d,e) except
$\theta_\mathbf{k}=0^{\circ}$.}
\end{figure}
Fig.\ref{eddycurrent}(a,b) give other two examples of this global
eddying currents in the same zigzag GNR as Fig.\ref{E56-61}, their
energies are slightly above the second and third van Hove peaks of
the conducting band, respectively. We can see that both of them have
pronounced vortices in the flat areas far beyond the ripple, where
the pseudo-magnetic field has certainly vanished. Generally, if the
ripple is slope relative to the GNR axis, the current distributions
within the energy range slightly above the bottom of a conducting
sub-band or below the top of a valence sub-band will occur
remarkable vortices in entire flat areas. These vortices appeared in
flat areas are rather exotic because there is no local responsible
deflection mechanism.

\begin{figure}[htb] \setlength{\unitlength}{1cm}
\centering
\begin{picture}(8, 3.5)(0,4.2)
\includegraphics[width=8cm, clip]{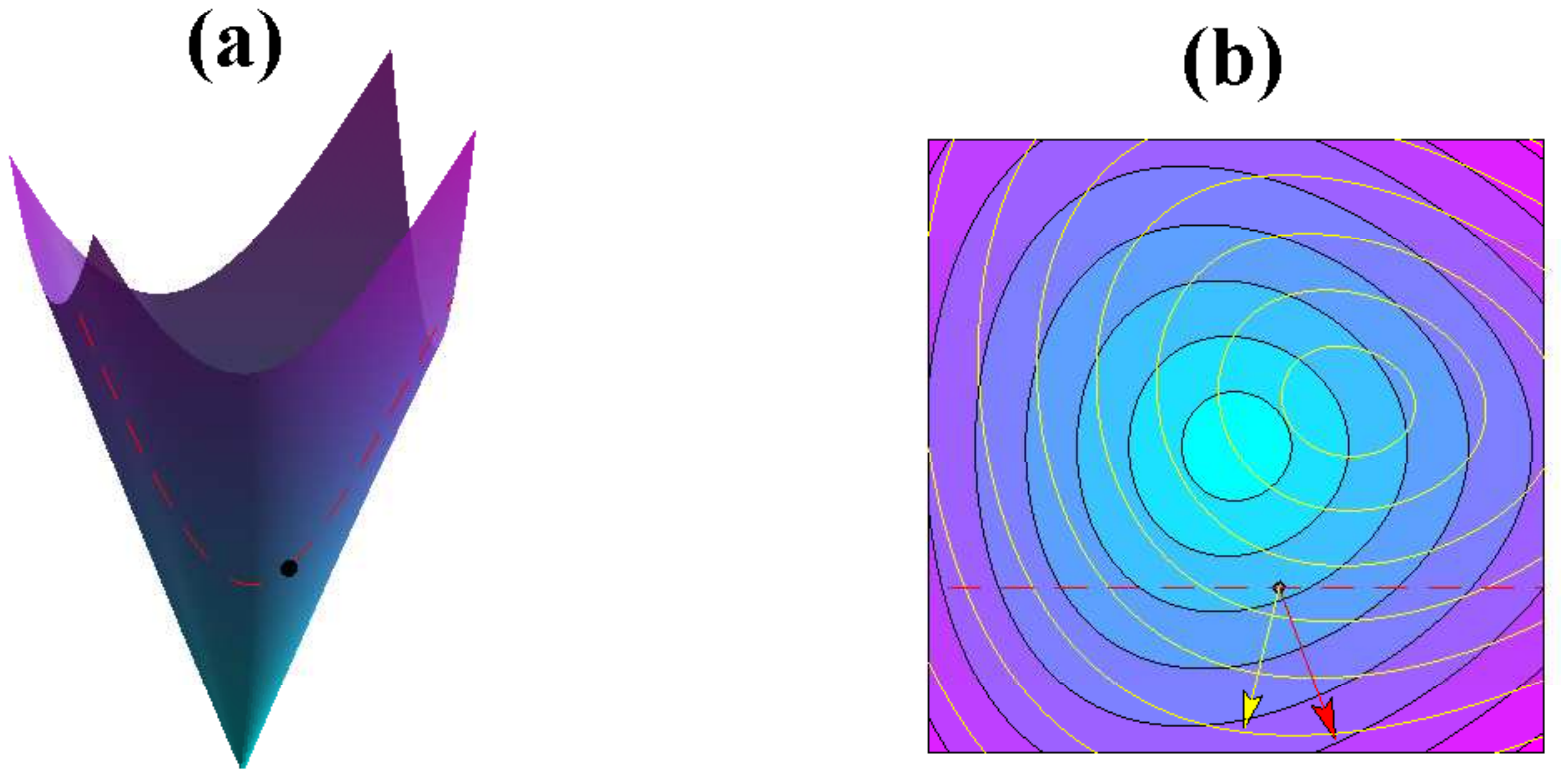}
\end{picture}
\begin{picture}(8, 3.2)(0,2.5)
\includegraphics[width=8cm, clip]{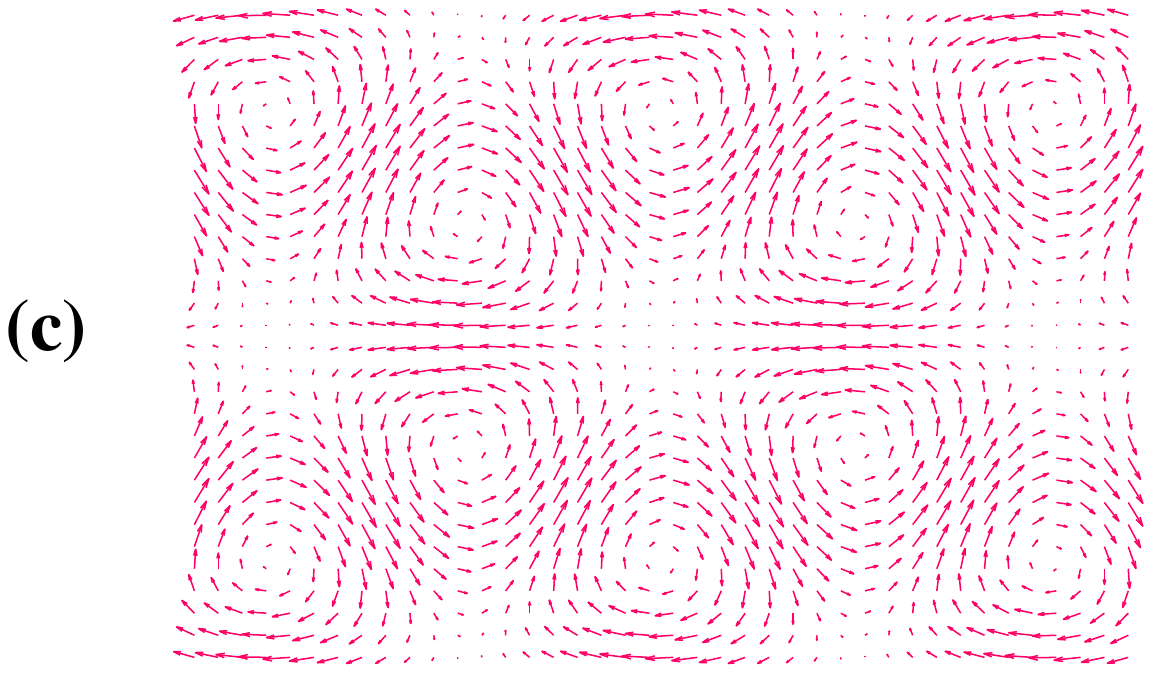}
\end{picture}
\caption{\label{band} (a) Conducting band of graphene near a Fermi
point, the red dashed line represents a sub-band of a GNR. The
velocity of states near the minimum of the sub-band (red dashed
line) can be easily reversed under a small variation of the energy
band, as shown in (b). (b) Contour map of energy bands for a perfect
(blue-purple) and a deformed (yellow) graphene. For the deformed
one, only two bonds are elongated respectively by $5.3\%$ and
$2.8\%$. The red(yellow) arrow represents the velocities of a state
in perfect(deformed) graphene. (c) Current distribution of three
plane waves.}\end{figure}

In order to explain the origin of these exotic vortices, we have to
notice two basic properties of the electronic states of GNRs under
perturbation. The first is that the electronic states will become
superposition of partial waves with approximate energies. The second
is that the velocity direction (forward or backward relative to
GNR's axis) of the states near the bottom of a conducting sub-band
(or the top of a valence sub-band) is unstable, because its velocity
$\partial E(\mathbf{k})/\partial k_x$ ($k_x$ is the momentum
component along the GNR axis) is very small, so its sign can be
easily changed by a small variation of $E(\mathbf{k})$, as
illustrated in Fig.\ref{band}(a,b).
 Therefore, if there is a nonuniform
deformation, the perturbed state corresponding to an originally
forward state in these energy ranges may contribute a backward flow
in the deformed regions. In particular, this local backward flow
demands the perturbed state to include backward partial waves in
flat areas in order to satisfy continuity condition at the
interfaces between deformed and flat areas, though there exists no
local responsible scattering mechanism. These partial waves of
different directions are always spatially superimposed due to the
edge reflection and will produce interference. Consequently, the
current density no longer equals to the sum of the currents of every
partial waves,
$\mathbf{j}(\mathbf{r})\neq\sum\langle\mathbf{k}|\hat{\mathbf{j}}(\mathbf{r})|\mathbf{k}\rangle$,
but must include interference terms
$\langle\mathbf{k}|\hat{\mathbf{j}}(\mathbf{r})|\mathbf{k}'\rangle$,
similar to the usual interference for probability density. These
kind of current interference patterns will be winding or eddying
flows in areas without any local deflection mechanism. As a simplest
example, Fig.\ref{band}(c) shows the current interference pattern of
three plane waves $\psi(\mathbf{r})=e^{i\mathbf{k}_1\cdot
\mathbf{r}}+e^{i\mathbf{k}_2\cdot \mathbf{r}}+e^{i\mathbf{k}_3\cdot
\mathbf{r}}$, where $\mathbf{k}_{1,2,3}$ have equal length while
their included angles are $120^{\circ}$. We can see that its current
distribution forms a very similar eddying pattern. Obviously, the
vortex scale of these eddy currents arising from the interference is
in proportional to the wavelengths of the partial waves. For the
eddy currents in rippled GNRs, this character can also be verified
by comparing Fig.\ref{E56-61}(a,d) and Fig.\ref{eddycurrent}(a,b).
We find that the vortices become smaller and smaller with the
increasing of energy owing to the linear dispersion relation
$E\propto k$. Conversely, the vortex scale will be very large in
wider GNRs because the energies of each corresponding step of the
conductance staircases will be smaller, it is only limited by the
electronic interference length.

According to this explanation, these global eddy currents would be a
rather ubiquitous effect in rippled GNRs, although their patters
depend on specific ripple configurations. Fig.\ref{eddycurrent}(c)
is another eddy current in an identical zigzag GNR in the same
energy range as Fig.\ref{E56-61}(b) but with different ripples,
which is a hyperbolic surface
\begin{eqnarray}\label{hypersurf}z(\mathbf{r})=\left\{\begin{array}{lc}h[1-f(\mathbf{r})],& f(\mathbf{r})\leq1 \\
0,&f(\mathbf{r})>1,\end{array}\right.\end{eqnarray}
where$f(\mathbf{r})=\sqrt{\left(\frac{\mathbf{\hat{e}}_1\cdot\mathbf{r}}{a_0}\right)^2
+\left(\frac{\mathbf{\hat{e}}_2\cdot \mathbf{r}}{b_0}\right)^2}$,
with $h=4a$, $a_0=15a$, $b_0=25a$,
$\mathbf{\hat{e}}_{1,2}=(\sqrt{2}/2,\mp \sqrt{2}/2)$(its maximum
bond elongation is about $3\%$). By comparing Fig.\ref{E56-61}(d)
and Fig.\ref{eddycurrent}(c), we can see that their current
distributions in flat regions are very similar although their ripple
are very different. Similar to this structural insensitivity of the
global eddying character, it is conceivable from the above
explanation that this vortical character will also not very
sensitive to the energy variation. Actually, the representation of
these eddying states in the LDOS curve is not a sharp peak like
quasi-bound states, but a broad and smooth one forming a piece of
continuous spectrum (Fig.\ref{LDOS}).

In addition, there are few special cases worth to be particularly
pointed out. The first case is zigzag GNRs with
$\theta_\mathbf{k}=0^{\circ}$. In this case the current lines remain
to be straight lines and no vortex occurs, because in this case the
$\mathbf{A}$ is parallel to $\mathbf{k}$ according
Eq.(\ref{potential}) or \cite{Yang}, so the $\nabla\times\mathbf{A}$
can be ignored; moreover, the incident current is along a symmetric
axis of the energy band, so the refraction also does not change its
direction. The second is zigzag GNRs with
$\theta_\mathbf{k}=\pm60^{\circ}$ or armchairs GNRs with
$\theta_\mathbf{k}=\pm30^{\circ}$. Similar to the first case, here
the $\nabla\times\mathbf{A}$ are also very small, however, there
exist apparent vortices, as shown in Fig.\ref{eddycurrent}(e),
because in these cases the anisotropic deformation of the energy
contours(see Fig.\ref{band}(b)) will result in similar backward
flow. The third one is armchair GNRs of
$\theta_\mathbf{k}=0^{\circ}$, although the ripple induce a strong
pseudo-magnetic field and results in apparent quasi-bound states
(Fig.\ref{LDOS}(f) and Fig.\ref{eddycurrent}(f)), but their currents
do not occur any vortex(Fig.\ref{eddycurrent}(g)), because the
direction of the incident current is the symmetric axis of two
pseudo-magnetic fields for two inequivalent Dirac points (mutually
symmetric) as well as the principle axis of the velocity tensor, so
the action of the pseudo-magnetic fields of two Dirac points will be
mutually canceled out and the refraction and dispersion due to the
velocity anisotropy also does not change the current direction.

In summary, the current flows near every step edge of the
conductance staircases of a GNR are rather unstable. They will
become eddy currents if there occurs a slope ripple. These eddy
currents can be divided into two classes. The first one are carried
by Landau-like states caused by the pseudo-magnetic field, these
states have slightly lower energies and their current distributions
form vortices only within the ripple region. In contrast, the second
one are carried by some special scattering states, which have
slightly higher energies and include backward partial waves in flat
areas. Consequently, they will form some global eddy currents due to
the interference of these partial waves. This global eddy current is
a manifestation of the non-locality of quantum interference effect.

\begin{acknowledgments} This work was supported by "the Fundamental Research Funds for the Central Universities" and
NSFC(Grant Nos. 10974027, 50832001).
\end{acknowledgments}

\end{document}